\documentclass[twocolumn,floats,showpacs,pre]{revtex4}
\usepackage[above]{placeins}
\usepackage{amsmath}
\usepackage{graphicx}

\begin{document}

\title{The Mystery of Superconductivity: Glue or No Glue?}
\author{X. Q. Huang$^{1,2}$}
\email{xqhuang@netra.nju.edu.cn} \affiliation{$^1$Department of
Telecommunications Engineering ICE, PLAUST, Nanjing 210016,
China \\
$^{2}$Department of Physics and National Laboratory of Solid State
Microstructure, Nanjing University, Nanjing 210093, China }
\date{\today}

\begin{abstract}
In this study,  a possible non-quasiparticle glue for
superconductivity of both conventional and unconventional
superconductors is explored in a pure electron picture. It is shown
clearly that the moving electrons due to the electromagnetic
interaction can self-organize into some quasi-one-dimensional
real-space charge stripes, which can further form some
thermodynamically stable vortex lattices with trigonal or tetragonal
symmetry. The relationships among the charge stripes, the Cooper
pairs and the Peierls phase transition are discussed. The suggested
mechanism (glue) of the superconductivity may be valid for the one-
and two-dimensional superconductors. We also argue that the highest
critical temperature of the doped superconductors is most likely to
be achieved around the Mott metal-insulator transition.

\end{abstract}


 \maketitle

\section{Introduction}

Since the discovery of superconductivity in mercury by K. Onnes in
1911\cite{onnes}, great efforts have been made to finding out how
and why it works. It has been widely accepted that the BCS
(Bardeen-Cooper-Schrieffer)\cite{bcs} successfully explained the
superconducting behavior in conventional superconductors by
predicting that the electrons near the Fermi surface can be `glued'
together in Cooper pairs by the attractive force of the
electron-phonon interaction. According to the microscopic BCS
theory, the maximum critical temperature ($T_{c}$) of
superconductors cannot exceed the McMillan limit of 39 K.

With the discovery of a family of cuprate-perovskite ceramic
materials known as high-temperature superconductors in 1986
\cite{bednorz,mkwu}, many theoretical condensed matter physicists
have started to doubt the reliability of the phonon-mediated BCS
theory \cite{anderson1}. As we know that the highest critical
temperature of cuprate superconductors ever recorded is
$HgBa_{2}Ca_{2}Cu_{3}O_{8}$ (under 30 GPa pressure) \cite{gao},
which has a critical temperature as high as 164 K. This indicates
that the gentle lattice vibrations (phonon-glue) may be not the
right candidate for high-temperature superconductivity. Recently,
the reliability of BCS theory has been further challenged by the new
iron arsenide superconductors with critical temperatures in excess
of 50 kelvin \cite{kamihara,zaren}. In fact, at high temperatures,
the vibration becomes so vigorous that it tends to break up the
electron pairs instead of binding them together \cite{anderson2}. So
what could possibly provide the \textquotedblleft
glue\textquotedblright\ for high temperature superconductivity?

As is well known, twenty-three years after the appearance of the
high-temperature superconductors, though more than 100,000 papers on
the materials have been published and many \textquotedblleft
glues\textquotedblright\ (for example, the magnetic resonance mode,
spin excitations and phonons) have been suggested, however,
scientists have been still debating the underlying physical
mechanism for this exotic phenomenon. Theorists have created a large
number of theoretical models for high-transition-temperature, as a
result, it makes the problem even more confusing. Just as Steven
Kivelson said, \textquotedblleft The theoretical problem is so hard
that there isn't an obvious criterion for right\textquotedblright\
\cite{cho}. In a recent paper, Anderson even questioned the
existence of any electron-pairing glues in cuprate superconductors
\cite{anderson2}. More recently, Pasupathy et al. \cite{pasupathy}
showed temperature-dependent scanning tunneling spectroscopy data
which has been believed to be strong evidence for the
\textquotedblleft no glue\textquotedblright\ superconducting
picture.

It is now quite clear that superconductivity can occur in a wide
variety of materials, including some simple elements (like niobium
and tantalum), various metallic alloys and organic materials. Thus,
it is not surprised to find that more and more materials with the
superconducting properties will be discovered in future. Most
theorists believe that new superconductors always reveal the need
for fresh mechanism and theoretical models. Moreover, they hope to
uncover the mystery of superconductivity simply through the
Hamiltonian, which has been discussed ad nauseum by now. But our
viewpoint is somewhat different from the these physicists. In our
opinion, if there exists only a few materials with the
superconductivity, it may be reasonable to expect that they have
different superconducting mechanisms. As so many materials with
superconductivity have been discovered, it becomes more clear that
the all superconducting phenomena should share an exactly the same
physical reason. Furthermore, the new mechanism for the
electron-pairing glue that gives rise to superconductivity should
not be established in Hamiltonian systems.

 In this paper, we will
present a non-quasiparticle \textquotedblleft
glue\textquotedblright\ which can naturally bring the moving
electrons together and condense them into some real-space
superconducting vortex lattice states.

\section{Can electrons attract each other?}

In spite of more than twenty years of long and difficult debates on
what causes high-temperature superconductivity, we insist that there
might just be a single and simple explanation for electron coupling
in various superconductors. Superconductivity, as a widespread
natural phenomenon should be governed by a unique and fundamental
deterministic law of nature.

\begin{figure}[tbp]
\begin{center}
\resizebox{0.9\columnwidth}{!}{
\includegraphics{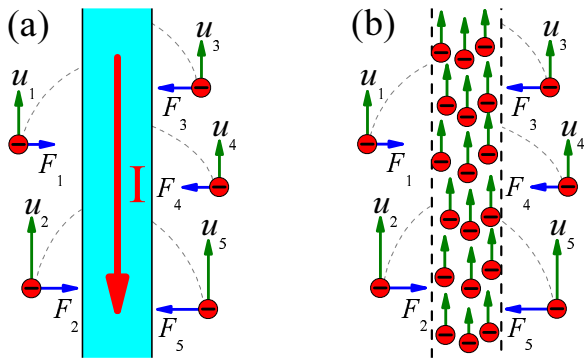}}
\end{center}
\caption{ Can electrons attract each other? (a) The moving electrons
are attracted to the electric current. (b) The electric current of
(a) is fully equivalent to the directional movement of electrons
inside the conductor, this implies that the electrons moving in the
same direction may mutually attract.  } \label{fig1}
\end{figure}

Normally, electrons repel each other according to Coulomb's law,
they are attracted only to protons or positive ions. However, it is
argued that electrons should attract each other in some particular
situations like superconductivity. Do electrons attract electrons?

As shown in Fig. \ref{fig1}, we have a new window on this critical
question of what holds electrons together. According to
electromagnetic theory, an electric current produces a magnetic
field which will exert force on the mobile electrons nearby, as
illustrated in Fig. \ref{fig1}(a). It is a common knowledge that it
is the directional movement of electrons which are responsible for
electric current in conductors such as wires, as shown in Fig.
\ref{fig1}(b). This figure implies a very important message that the
electrons moving in the same direction may mutually attract, rather
than mutually repel in their resting state. It seems likely that
this fundamental physical fact or property has been overlooked by
the researchers of condensed matter physics. This scenarios of Fig.
\ref{fig1}(b) provides a natural glue (without the concept of
quasiparticle) which can bring the moving electrons together, a more
detailed discussion will be given in next section.

\section{The electromagnetic interaction induced self-organization of moving electrons}

\begin{figure}[tp]
\begin{center}
\resizebox{0.95\columnwidth}{!}{
\includegraphics{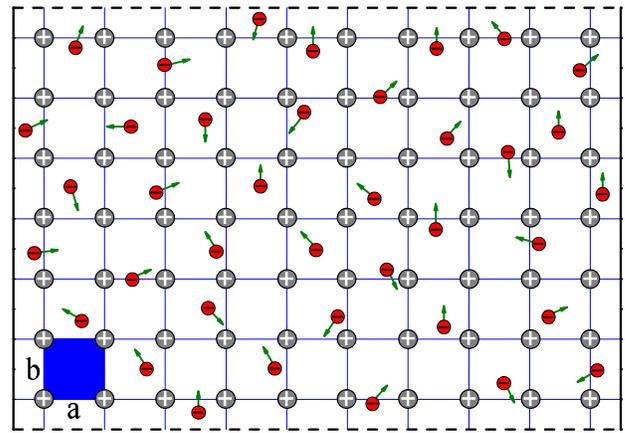}}
\end{center}
\caption{One plane in a superconducting material, which contains
positively charged lattice structure and a certain concentration of
charge carriers (electrons). } \label{fig2}
\end{figure}

From the view of crystallography, all superconducting materials can
be simply depicted in Fig. \ref{fig2}, which contains two basic
elements: (1) the lattice structure of  positive charge (ions), and
the carriers of negative charge (electrons). Here, we may raise one
most essential question: what is the fundamental difference between
the superconducting materials and the non-superconducting materials?
The answer is simple and definite: to be a superconductor, the
materials should include an appropriate carrier number, or with an
appropriate carrier concentration (not too high, not too low). Of
course, in order to obtain a higher superconducting temperature, we
will show that some matching conditions between the carrier
concentration and the lattice structure should be naturally
satisfied.

\begin{figure}[tp]
\begin{center}
\resizebox{1\columnwidth}{!}{
\includegraphics{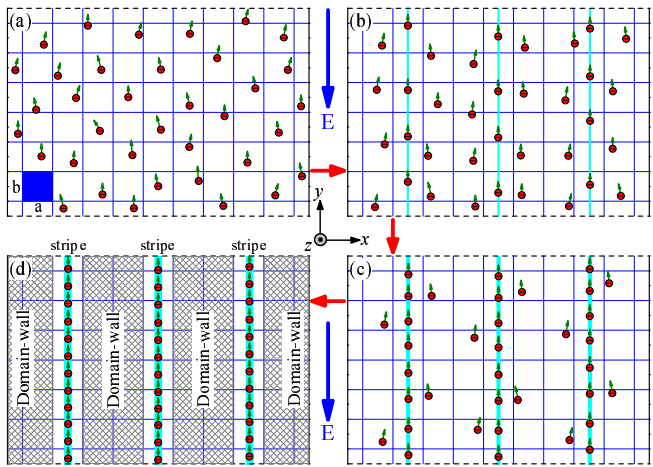}}
\end{center}
\caption{The external electric field induced self-organization of
the charge carriers. (a) An overall movement of the electrons in
$y$-direction, (b) the appearance of the blurred charge rivers, (c)
more electrons join the charge rivers, (d) the moving electrons are
finally condensed into some quasi-one-dimensional charge stripes (or
superconducting vortex lines) separated by the domain wall of of the
positive ions.}\label{fig3}
\end{figure}

It is well known that the application of an external electric field
(in $-y$ direction) on a material can cause an overall movement of
the charge carriers (electrons) in $y$ direction, see Fig.
\ref{fig3}(a). Under the conditions of low temperature and low
carrier concentration, it seems likely that new physical phenomena
will emerge, as shown in Figs. \ref{fig3}(b)-(d). As time goes by,
the electrons of the directional movement will gradually gather
together and self-organize into some highly ordered charge
structures due to the magnetic field forces as illustrated From Fig.
\ref{fig3}(b) to Fig. \ref{fig3}(d). Finally, these moving electrons
will condensed into some quasi-one-dimensional charge stripes
(vortex lines), or \textquotedblleft charge
rivers\textquotedblright\ \cite{emery,Tranquada}. In this case, the
corresponding superconductor exhibits a peculiar form of real-space
phase separation, these \textquotedblleft charge
rivers\textquotedblright\ are formed spontaneously and segregated by
the domain walls of the positive ions, as shown in Figs.
\ref{fig3}(d). Based on the energy minimization principle, to be a
stable superconducting phase, these charge stripes (vortex lines)
must be arranged in periodic arrays in the superconducting plane,
also see Figs. \ref{fig3}(d).

\begin{figure}[tbp]
\begin{center}
\resizebox{1\columnwidth}{!}{
\includegraphics{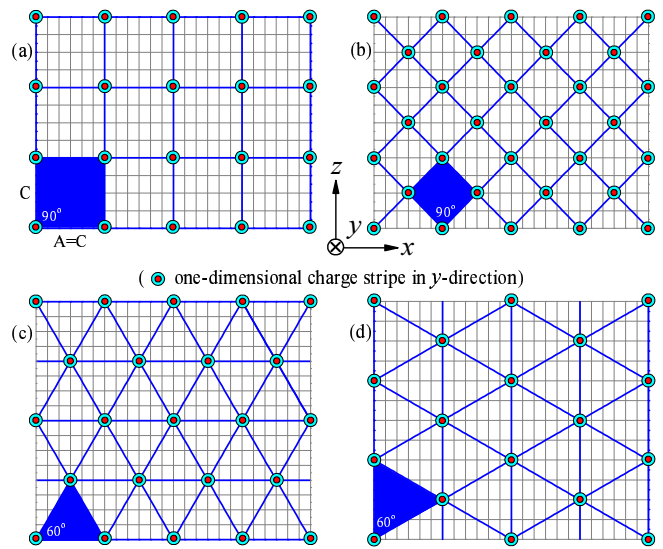}}
\end{center}
\caption{To achieve the highest superconducting transition
temperature, the superconducting vortex lattices should be in the
following four stable structures, (a) and (b) the vortex lattices
with tetragonal symmetry, while (c) and (d) having the  trigonal
symmetry. } \label{fig4}
\end{figure}

The self-organization of the moving electrons, which I find to be
one of the most interesting and important concepts in modern
physics. Obviously, the self-organization picture provides a vivid
description of the one- and two-dimensional superconductivity [see
Fig. \ref{fig3}(d)]. For the three-dimensional bulk superconductors,
the quasi-one-dimensional superconducting vortex lines will
reorganize into a vortex lattice. We argue that the superconductor
with a maximum critical temperature is that at which a uniform
distribution of vortex lines in the plane perpendicular to the
stripes. In this sense,  the low-temperature tetragonal phases
[Figs. \ref{fig4}(a)-(b)] and the simple hexagonal phases [Figs.
\ref{fig4}(c)-(d)] might be the ideal candidates for the most stable
vortex lattices of the charge stripes.

Here it is worth mentioning that, for the doped superconductors (for
example, the cuprate and iron pnictide superconductors), the carrier
concentration can be adjusted within a relatively wide range and
this in turn will influence the structure of the superconducting
vortex lattices of Figs. \ref{fig4} and the superconducting critical
temperature. According to our theory, with decreasing the carrier
concentration, the stripe-stripe interactions inside the vortex
lattice can be significantly reduced and hence enhance the material
superconducting transition temperature. Based on this idea, we can
argue that the highest $T_{c}$ of the doped superconductors may be
achieved around the Mott metal-insulator transition, where the
superconducting vortex lattice's energy takes the smallest value. (A
more detail discussions will be presented in presented in another
paper.)

\section{Peierls phase transition and Cooper pairs}

In the framework of the self-organized of the moving electrons and
the vortex lattices of Fig. \ref{fig4}, what factors can affect the
superconducting transition temperature?

\begin{figure}[bp]
\begin{center}
\resizebox{1\columnwidth}{!}{
\includegraphics{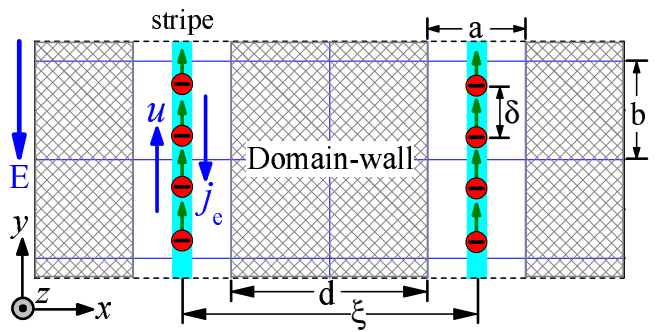}}
\end{center}
\caption{The required  physical parameters for the  describing of
the superconducting vortex lines, where $u$ is the electron-transfer
velocity and  $j_{e}$ is the current density of a single vortex
line. The other parameters are explained in the text.}
\label{fig5}\end{figure}

Figure \ref{fig5} shows an area of superconducting plane, where $\xi
$ is the stripe-stripe separation, $d$ is the width of domain-wall,
$a$ and $b$ are the lattice constants, and $\delta$ is the
electron-electron distance within one vortex line. In general, all
of these structure parameters can influence the critical temperature
of the corresponding superconductor. Qualitatively, for a relatively
wide domain-wall (or a larger stripe-stripe separation), the
stripe-stripe interactions will be greatly reduced, and consequently
enhance the stability of the superconducting state which in turn
improve the superconducting transition temperature of the
superconductor. Hence, the high-temperature superconducting
materials typically have a very low carrier concentration (for
example, the cuprate superconductors), while the conventional
superconductors have a relatively high  carrier concentration.
Moreover, the materials with an exceptionally high carrier
concentration (the electrons are too crowded to form the order
stripes.) may not be superconductors at any low temperatures, such
as gold, silver and copper, the most common good conductor of
electricity. To maintain a more stable charge river (vortex line),
the mobile electrons must be effectively confined in some
quasi-one-dimensional spaces [say the cyan lines in Fig.
\ref{fig3}(d)], usually, a small lattice constant $a$ and a thick
domain wall $d$ are conducive to the stability of the charge river
and a higher $T_{c}$ superconducting state.

\begin{figure}[tbp]
\begin{center}
\resizebox{1\columnwidth}{!}{
\includegraphics{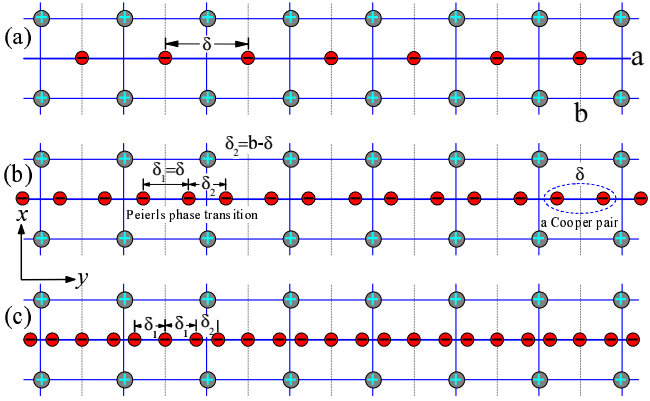}}
\end{center}
\caption{(a) Non-pairing superconducting vortex line may survival in
the superconductors with a small lattice constant $b$. (b) The
superconducting vortex lines are most likely in a Peierls chain with
the electron-electron separations $\delta$ and $b-\delta$, in this
special case, the Cooper pairs can naturally form inside each
plaquette alone the vortex lines. (c) An unstable vortex line that
may be easily destroyed by the strong electron-electron interactions
among the crowded electrons.} \label{fig6}
\end{figure}

As an approximate description, we use a single parameter $\delta$ to
characterize a superconducting vortex line in Fig. \ref{fig5}. It
should be pointed out that the formation of charge stripe is
generally attributed to the competition between the short-range
electron-electron static electric repulsion and the long-range
dynamic magnetic attraction of Fig. \ref{fig1}(b). Consequently,
there exists an optimal electron-electron separation within the
vortex line, We will discuss in next paper that, to be a
superconducting vortex line, the corresponding electron-electron
separation should lay in the range of $1.4\sim 1.8\mathring{A}$. In
addition, we must emphasize that the one parameter's description
(see Fig. \ref{fig5}) of the vortex line is not accurate, for a real
superconductor, the electron-electron distance inside one vortex
line is modulated by lattice structure of the superconductor.

Further, we consider the effects of the lattice structure of the
superconductor on the formation of the one-dimensional vortex lines.
As shown in Fig. \ref{fig6}, the vortex lines in the quasi-static
state may have different structures. Fig. \ref{fig6}(a) shows a
periodic vortex line with the electron-electron separation $\delta$
equals to the lattice constant $b$, in the case of small lattice
constant, such vortex lines may exhibit the superconductivity
phenomenon. This may be considered as a non-pairing mechanism of
superconductivity. However, for a large lattice constant $b$, the
moving vortex lines will be very unstable due to a rather weak
magnetic attraction depicted in Fig. \ref{fig1}. For most materials,
the lattice constant $b$ usually ranges between $3\mathring{A}$ to
$4\mathring{A}$, which is about two times as large as the optimal
electron-electron separation ($\sim 1.5\mathring{A}$). This implies
that there are (average) two electrons (Cooper pair) inside one
plaquette along one superconducting vortex line, as shown in Fig.
\ref{fig6}(b) which can be considered as the lattice structure
induced Peierls phase transition. If there are more than three
electrons inside one plaquette [see Fig. \ref{fig6}(c)],  the
electrons are too crowded inside the vortex lines and the
electron-electron static electric repulsions are strong enough to
break up the vortex lines.

\begin{figure}[tbp]
\begin{center}
\resizebox{1\columnwidth}{!}{
\includegraphics{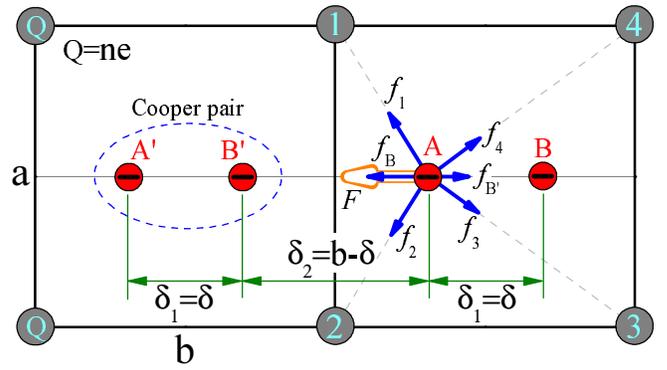}}
\end{center}
\caption{A picture of detailed illustration of electron-electron and
ion-electron interactions inside a vortex line. } \label{fig7}
\end{figure}

In the following, we will focus our attention on the formation of
the Cooper pair in Fig. \ref{fig6}(b). Along one superconducting
vortex line, the electrons are dimerized into Cooper pairs with a
spacing of $\delta$, as shown in Fig. \ref{fig7}. So the electrons
moving inside the vortex lines are in the energy minimum Peierls
chains. For the purpose of a simplified case, we consider only the
nearest-neighbor electron-electron and ion-electron interactions.
Based on Figure \ref{fig7}, the nearest-neighbor electron-electron
interactions on \textbf{electron A} can be expressed as:

\begin{eqnarray}
f_{B} &=&\frac{e^{2}}{4\pi \varepsilon _{0}\delta ^{2}},  \label{fB} \\
f_{B^{\prime }}&=&-\frac{e^{2}}{4\pi \varepsilon _{0}(b-\delta
)^{2}}, \label{fB1}
\end{eqnarray}%

If for each lattice ion carrying a positive charge $Q$ , we can get
the nearest-neighbor ion-electron interactions on \textbf{electron
A} as

\begin{equation}
f_{1}+f_{2}=\frac{2Qe(b-\delta )}{\pi \varepsilon
_{0}[a^{2}+(b-\delta )^{2}]^{3/2}},  \label{f12}
\end{equation}%
and
\begin{equation}
f_{3}+f_{4}=-\frac{2Qe(b+\delta )}{\pi \varepsilon
_{0}[a^{2}+(b+\delta )^{2}]^{3/2}}.  \label{f34}
\end{equation}%

Now we have a general formula of the total confinement force $F$ applied to the \textbf{electron A} (or \textbf{B}) of the Cooper pair as%

\begin{equation}
F=f_{B}+f_{B^{\prime }}+f_{1}+f_{2}+f_{3}+f_{4}.  \label{ftotal}
\end{equation}%

\begin{figure}[tp]
\begin{center}
\resizebox{1\columnwidth}{!}{
\includegraphics{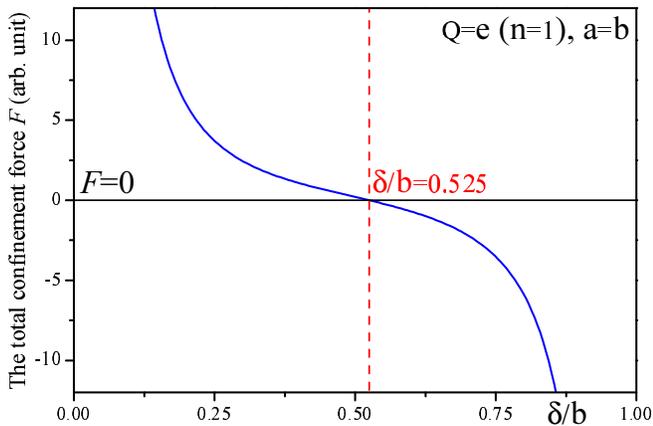}}
\end{center}
\caption{Analytical total confinement force $F$ versus $\delta/b$
inside one plaquette. Under the special conditions of $a=b$ and
$Q=e$, a quasi-static Cooper pair may exist in the plaquette with a
electron-electron separation $\delta=0.525b$. } \label{fig8}
\end{figure}

Physically, when $F$  is equal to zero, it indicates a completely
suppression of the Coulomb repulsion between two electrons. As a
consequence, the electrons will be in the energy minimum bound
state. Based on the analytical expressions (\ref{fB})$-$(\ref{f34}),
we draw in Fig. \ref{fig8} the confinement force $F$ versus
$\delta/b$ under the conditions $Q=e$ (or $n=1$)and $a=b$. This
figure reveals one important fact: the combination of the
ion-electron and electron-electron interactions can lead to the
well-known Peierls phase in the superconducting vortex lines, where
there are two electron-electron separations of
$\delta_{1}=\delta=0.525b$ and  $\delta_{2}=b-\delta=0.475b$.

\section{Concluding remarks}

We have proposed the self-organized picture of vortex lines due to
the electromagnetic interaction of the mobility electrons. It has
been shown clearly that the electrons moving in the same direction
may mutually attract, rather than mutually repel in their resting
state. In our approach, the microscopic scenario for the
superconductivity can be considered as a \textquotedblleft no
glue\textquotedblright\ superconducting picture because no any
quasiparticles are involved in the suggested mechanism. This no
quasiparticle characteristic implies that the proposed scheme
represents an unified interpretation of the superconductivity
phenomena of any kind of superconductors. We think that the
suggested real space self-organized mechanism of the charge carrier
may finally shed light on the mysteries of superconductivity.
Furthermore, our researches also reveal that the Peierls phase
transition is induced by the ion-electron interactions, rather than
spontaneous generation. We have argued that the highest $T_{c}$ of
the doped superconductors may be achieved around the Mott
metal-insulator transition, where the suggested real-space
superconducting vortex lattice is in its minimum energy state.

\end{document}